**Socio-meteorology: flood prediction, social preparedness, and cry wolf effects**

Yohei Sawada[1], Rin Kanai[2], and Hitomu Kotani[1,3]

[1] Institute of Engineering Innovation, the University of Tokyo, Tokyo, Japan

[2] Department of Civil Engineering, the University of Tokyo, Tokyo, Japan

[3] Department of Urban Management, Kyoto University, Kyoto, Japan

Corresponding author: Y. Sawada, Institute of Engineering Innovation, the University of Tokyo, Tokyo, Japan, 2-11-6, Yayoi, Bunkyo-ku, Tokyo, Japan, yohei.sawada@sogo.t.u-tokyo.ac.jp

**Abstract**

To improve the efficiency of flood early warning systems (FEWS), it is important to understand the interactions between natural and social systems. The high level of trust in authorities and experts is necessary to improve the likeliness of individuals to take preparedness actions responding to warnings. Despite a lot of efforts to develop the dynamic model of human and water in socio-hydrology, no socio-hydrological models explicitly simulate social collective trust in FEWS. Here we develop the stylized model to simulate the interactions of flood, social collective memory, social collective trust in FEWS, and preparedness actions responding to warnings by extending the existing socio-hydrological model. We realistically simulate the cry wolf effect, in which many false alarms undermine the credibility of the early warning systems and make it difficult to induce preparedness actions. We found (1) considering the dynamics of social collective trust in FEWS is more important in the technological society with infrequent flood events than in the green society with frequent flood events; (2) as the natural scientific skill to predict flood events is improved, the efficiency of FEWS gets more sensitive to the behavior of social collective trust, so that forecasters need to determine their warning threshold by considering the social aspects.

**1. Introduction**

The number of severe flood events is expected to increase in many regions due to climate change (Hirabayashi et al. 2013, 2021). Based on the advances of weather forecasting (e.g., Bauer et al. 2015; Miyoshi et al. 2016; Sawada et al. 2019) and hydrodynamic modeling (e.g., Yamazaki et al. 2011; Trigg et al. 2016), Flood Early Warning Systems (FEWS) have become the promising tool to efficiently mitigate the damage of severe floods. However, to maximize the potential of FEWS, it is crucially important to understand the interactions between flood and social systems. The likeliness of individuals to take preparedness actions responding to flood warnings strongly depends on the individual's risk perception which is controlled by the complex interaction between natural hazards and stakeholders (Wachinger et al. 2013).

In the literature of weather forecasting, the "cry wolf effect" has been intensively investigated as an important interaction between weather prediction and social systems. In Aesop's fable, the "The Boy who Cried Wolf", a young boy repeatedly tricks neighboring villagers into believing that a wolf is attacking the sheep. When a wolf actually appears and the young boy seriously calls for help, the villagers no longer trust the warning and fail to protect their sheep. Many false alarms undermine the credibility of the early warning systems. The cry wolf effect on mitigation and protection actions against meteorological disasters has been investigated in economics, sociology, and psychology. Simmons and Sutter (2009) performed econometric analysis of a disaster database and revealed that tornadoes that occurred in areas with higher false alarm ratio killed and injured more people. Ripberger et al. (2015) performed a web-based questionnaire survey and revealed that subjective perceptions of warning system's accuracy are systematically related to trust in a weather agency and stated responses to warnings. Trainor et al. (2015) performed large-scale telephone interviews and revealed the significant relationship between actual false alarm ratio and behavioral responses to tornado warnings. They also found that there is a wide variation in public definition of false alarms and actual false alarm ratio does not predict perception of false alarm ratio, which illustrated the significant complexity associated with the analysis of false alarms. Although Trainor et al. (2015) could not find the significant relationship between perceived false alarm ratio and responses to warnings, Jauernic and van den Broeke (2017) revealed that the odds of students initialing sheltering decreases nearly 1% for every 1% increase in perceived false alarm ratio based on their online questionnaire survey of 640 undergraduate students. While these previous works supported the cry wolf effect as an important factor to be considered for the design of warning systems, many existing studies discussed the myth of cry wolf effects implying that they do not exist. For example, LeClerc and Joslyn (2015) performed a psychological experiment in which participants decided whether to apply salt brine to a town's roads to prevent icing according to weather forecasting. In their experiment, the effects of false alarms are so small that they found no evidence suggesting lowering false alarm ratio significantly increases compliance with weather warnings. Lim et al. (2019) performed an online questionnaire survey and found no significant relationship between actual false alarm ratio and responses to warnings. In addition, they found that the increase of perceived false alarm ratio enhanced protective behavior, which contradicted the other works. Although the existence of the cry wolf effect is still debatable, the warning threshold of the actual weather warning systems can be justified only if the cry wolf effect is considered (Roulston and Smith 2003). It is crucially important to understand the effect of false alarms on behavioral responses to warnings to design efficient weather warning systems.

Socio-hydrology is an emerging research field to contribute to understanding the interactions between flood and social systems (Sivapalan et al. 2012, 2014; Di

Baldassarre et al. 2019). The primary approach of socio-hydrology is to develop the dynamic model of water and human. Many socio-hydrological models used social preparedness as a key driver of human-water interactions (e.g., Di Baldassarre et al. 2013; Viglione et al. 2014; Ciullo et al. 2017; Yu et al. 2017; Albertini et al. 2020). The pioneering work of Girons Lopez et al. (2017) revealed the effect of social preparedness on the efficiency of FEWS. Their main finding is that social preparedness is an important factor for flood loss mitigation especially when the accuracy of the forecasting system is limited. However, to our best knowledge, the existing socio-hydrological models simulated social preparedness as a function of social collective memory or personal experience of past disasters, and they considered no effect of trust in authorities and experts. Therefore, the cry wolf effect cannot be analyzed in the existing models. The systematic review of Wachinger et al (2013) indicated that both personal experience of past disasters and trust in authorities and experts have the substantial impact on risk perception. It is crucially important to include the social collective trust in FEWS in the socio-hydrological model to improve the design of FEWS considering social system dynamics.

The aim of this study is to develop the stylized model of the responses of social systems to FEWS as the simple extension of Girons Lopez et al. (2017). By modeling the dynamics of social collective trust in FEWS as a function of the recent success and failure of the forecasting system, we realistically simulate the cry wolf effect. By analyzing our newly developed model, we provide useful implication to maximize the potential of FEWS considering social system dynamics.

## 2. Model

Here we slightly modified the model proposed by Girons Lopez et al. (2017). For brevity, the detailed explanation of equations shared with Girons Lopez et al. (2017) is omitted in this paper. See Gironz Lopez et al. (2017) and references therein for the complete description.

A synthetic time series of river discharge is generated. Following Girons Lopez et al. (2017), a simple bivariate gamma distribution, $\Gamma$, is used:

$Q \sim \Gamma(\kappa_c, \theta_c)$   (1)

where Q is maximum annual flow. The bivariate gamma distribution is characterized by shape $\kappa_c$ and scale $\theta_c$.

This maximum annual flow, Q, is forecasted. In our model, the ensemble flood forecasting system (e.g., Cloke and Hornberger 2009) is installed and the probabilistic forecast can be issued. The forecast probability distribution, $F$, is calculated by the following:

$$F \sim N(Q + N(\mu_m, \sigma_m^2), N(\mu_v, \sigma_v^2)) \quad (2)$$

where $N(.)$ is the Gaussian distribution, $N(\mu_m, \sigma_m^2)$ controls the prediction accuracy, and $N(\mu_v, \sigma_v^2)$ controls the prediction precision. While Girons Lopez et al. (2017) changes $\mu_m$ in their simulation, we set $\mu_m = 0$ assuming the forecast is unbiased. While Girons Lopez et al. (2017) used the bivariate gamma distribution to model the prediction precision, we used the Gaussian distribution to make it easier to interpret results.

There is a damage threshold, $\delta$, which is the proxy of levee height. When $Q > \delta$, flood occurs. The forecast system calculates the probability of river discharge exceeding $\delta$ and issues a warning if this probability of exceedance, $P$, is larger than a predefined probability threshold, $\pi$. Table 1 summaries four different outcomes of forecasting: true positive, false positive, false negative, and true negative. When forecasters choose lower $\pi$, they issue many warnings with low forecasted probability of flooding, which inevitably increases false alarms. When forecasters choose higher $\pi$, they can reduce the number of false alarms by issuing the smaller number of warnings, which inevitably increases missed events.

Based on these four different outcomes shown in Table 1, damages and costs are calculated. Flood damage is assumed to be negligible when river discharge is smaller than a damage threshold (i.e. $Q < \delta$). When $Q \geq \delta$, the damage function is defined as a simple exponential function, which is often used in the socio-hydrological literature (e.g., Di Baldassarre et al. 2013):

$$D_Q = \begin{cases} 0 & (Q < \delta) \\ 1 - e^{-\frac{Q-\delta}{\beta}} & (Q \geq \delta) \end{cases} \quad (3)$$

where $D_Q$ is damage, $\beta$ is a model parameter. If a flood event is successfully forecasted and a warning is issued (i.e. $P \geq \pi$), this damage is mitigated by preparedness actions. How much damage can be mitigated depends on social preparedness, $P_r$. The mitigated damage (called residual damage in Girons Lopez et al. (2017)), $D_r$, is calculated by the following:

$$D_r = D_Q e^{-P_r \ln(\frac{1}{\alpha_0})} \quad (4)$$

where $\alpha_0$ is a model parameter which determines the minimum possible damage. In summary, the flood damage, $D$, can be described by equation (5):

$$D = \begin{cases} 0 & (Q < \delta) \\ 1 - e^{-\frac{Q-\delta}{\beta}} & (Q \geq \delta \text{ and } P < \pi) \\ \left(1 - e^{-\frac{Q-\delta}{\beta}}\right) e^{-P_r \ln\left(\frac{1}{\alpha_0}\right)} & (Q \geq \delta \text{ and } P \geq \pi) \end{cases} \quad (5)$$

Whenever a warning is issued, the cost, $C$, arises from mitigation and protection actions. Following Girons Lopez et al. (2017), we assumed that the cost is calculated by:

$$C = \begin{cases} 0 & P < \pi \\ \eta Q & P \geq \pi \end{cases} \quad (6)$$

where $\eta$ is a parameter.

The dynamics of social preparedness, $P_r$, in this study is different from Girons Lopez et al. (2017). We assumed that the social preparedness consists of social collective memory and social collective trust in FEWS:

$$P_r(t) = \gamma E(t) + (1-\gamma) T(t) \quad (7)$$

where $E(t)$ and $T(t)$ are social collective memory and social collective trust in FEWS at time $t$, respectively. $\gamma$ is a model parameter that weights $E(t)$ and $T(t)$. In many socio-hydrological models, social collective memory is driven by the recency of past flood experience. Following Girons Lopez et al. (2017), the dynamics of social collective memory is described by the following:

$$E(t+1) = \begin{cases} E(t) - \lambda E(t) & (D = 0) \\ E(t) + \chi D & (D > 0) \end{cases} \quad (8)$$

where $\lambda$ and $\chi$ are model parameters.

We assumed that social collective trust in FEWS is affected by the recent accuracy of FEWS. Previous studies pointed out that the recent forecast accuracy and false alarm ratio affected the performance of preparedness actions (Simmons and Sutter 2009; Trainor et al. 2015; Ripberger et al. 2015; Jauernic and van den Broeke 2017). It is reasonable to assume that trust in FEWS increases (decreases) when prediction succeeds (fails) (Wachinger et al. 2013). We propose the following simple equation to describe the dynamics of social collective trust in FEWS:

$$T(t+1) = \begin{cases} T(t) & for\ true\ negative \\ T(t) + \tau_{TP} & for\ true\ positive \\ T(t) - \tau_{FN} & for\ false\ negative \\ T(t) - \tau_{FP} & for\ false\ positive \end{cases} \quad (9)$$

where $\tau_{TP}$, $\tau_{FN}$, and $\tau_{FP}$, are positive parameters. By changing the value of these parameters, we can change the sensitivity of social collective trust in FEWS to the accuracy of FEWS. We will analyze the behavior of our model associated with several different combinations of these three parameters.

In our equations (7-9), we can consider both social collective memory and social collective trust to analyze behavioral responses to warnings. For instance, please assume that a severe flood occurs and substantially damages a community, and this flood events cannot be predicted. In this case, social collective memory increases due to the large damage (equation (8)). This increase of social collective memory $E(t)$ contributes to increasing social preparedness towards the next severe flood events (equation (7)). However, the failure of predicting this flood events decreases social collective trust in FEWS and authorities related to warning systems (equation (9)), which negatively impacts to the capability of a community to deal with the next flood events by decreasing social preparedness (equation (7)).

If social preparedness is determined only by social collective memory as Girons Lopez et al (2017) proposed, social preparedness constantly decreases and goes to 0 when no floods occur for a long while. In our proposed model, high social collective trust in FEWS can maintain the high level of social preparedness even if a community completely loses past flood experiences (equation (7)). However, if a weather agency repeatedly issues false alarms, social collective trust in FEWS decreases (equation (9)), which negatively impacts to social preparedness (equation (7)). Therefore, the dynamics of social preparedness in our proposed model is greatly different from Girons Lopez et al. (2017).

Many of the model parameters are fixed in our analysis. Table 2 summarizes the description and values of the fixed parameters. Some parameters are changed in our analysis to check their sensitivity to the performance of FEWS. Those parameters are explained in the next section.

## 3. Experiment design

### 3.1. Metrices

We used several metrics to evaluate the performance of FEWS. The purpose of FEWS is to reduce the total loss ($D + C$). We used the relative loss as Girons Lopez et al. (2017) did. The relative loss, $L_r$, is defined by equation (10):

$$L_r = \frac{L_{FEWS}}{L_{noFEWS}} \quad (10)$$

We performed the long-term (1000-year) numerical simulation by solving equations (1-9) and calculated the total loss, $L_{FEWS}$. We also performed the simulation without FEWS, in which flood damage is always calculated by equation (3) and $D$ is always equal to $D_Q$. The total loss of this additional simulation is defined as $L_{noFEWS}$. The relative loss measures the efficiency of FEWS.

In addition to relative loss, we used hit rate, false alarm ratio, and threat score to evaluate the prediction accuracy, which is not related to social system dynamics. They are defined by equations (11-13):

$$hit\ rate = \frac{O_{TP}}{O_{TP}+O_{FN}} \quad (11)$$

$$false\ alarm\ ratio = \frac{O_{FP}}{O_{FP}+O_{TP}} \quad (12)$$

$$threat\ score = \frac{O_{TP}}{O_{TP}+O_{FP}+O_{FN}} \quad (13)$$

where $O_{TP}$, $O_{FN}$, and $O_{FP}$ are the total number of true positive, false negative, and false positive events, respectively.

### 3.2. Simulation Settings

We firstly compared the original model proposed by Girons Lopez et al. (2017) with our modified model. When we set $\gamma = 1$ in equation (7), our model reduces to Girons Lopez et al. (2017) since we have no contributions of social collective trust in FEWS to social preparedness. In this paper, this original model is hereafter called the GL model. On the other hand, when we set $\gamma = 0.5$ in equation (7), our model considers both social collective memory and social collective trust in FEWS with same weights to calculate social preparedness. This new model is hereafter called the SKK model.

In the experiment 1, the timeseries of state variables of the two models are compared to demonstrate how differently the SKK and GL models work. The parameter variables in the experiment 1 are shown in Table 3.

We mainly focused on the relationship between relative loss and a predefined probability threshold, $\pi$. This warning threshold is important for forecasters to determine whether they require general citizens to take preparedness actions. In the experiment 2, we used the same damage threshold, $\delta$, as Girons Lopez et al (2017) and compared the relationship between relative loss and predefined warning thresholds in the GL model with that in the SKK model under the different prediction skills and the cost parameter $\eta$. The settings of the parameters in the experiment 2 can be found in Table 4.

In the experiment 3, we also compared the GL and SKK models under different damage thresholds, $\delta$. In socio-hydrology, previous works focused on the difference between "green" and "technological" society (Ciullo et al. 2017). In green society, the flood protection level is so low that many flood events occur, which increases social collective memory of flood events. In technological society, the flood protection level is high. Since flood events occur less frequently in the technological society, the high level of social collective memory cannot be maintained. By changing the damage threshold, we analyzed how differently the GL and SKK models behave in the different society. The settings of the parameters in the experiment 3 can be found in Table 5.

In the experiment 4, we analyzed only the SKK model. The primary purpose of this experiment 4 is to find the optimal predefined warning threshold, which minimizes relative loss, in not only different society and prediction accuracy but also different combinations of parameters related to the dynamics of social collective trust in FEWS (i.e., $\tau_{TP}, \tau_{FN}$, and, $\tau_{FP}$ in equation (9)). The settings of the parameters in the experiment 4 can be found in Table 6.

In experiments 2–4, we performed the 250-member Monte-Carlo simulation by randomly perturbing a predefined probability threshold, $\pi$, and the initial conditions of social collective memory and social collective trust in FEWS. We analyzed the sensitivity of the efficiency of FEWS to predefined warning thresholds.

## 4. Results

Figure 1 shows the time series of social preparedness of the GL and SKK models in the experiment 1 (see Table 3). In the GL model (Figure 1a), social preparedness (black line) increases when flood occurs (red and green bars) and is not affected by false alarms (blue bars). In the SKK model (Figure 1b), false alarms negatively impact social preparedness by reducing social collective trust in FEWS (pink line). From $t = 430$ to $t = 440$, consecutive false alarms substantially decrease social collective trust in FEWS and social preparedness, so that the damage of severe flood at $t = 452$ in the SKK model is larger than that in the GL model despite the accurate warning being issued. It is the cry wolf effect.

Figure 2a shows the relationship between relative loss and predefined warning thresholds simulated by the GL model in the experiment 2 (see Table 4). We firstly assumed that there is no cost of the mitigation and protection action and is the relatively accurate prediction system (the experiment 2.1; see Table 4). In this case, FEWS can minimize the relative loss with the extremely small predefined warning thresholds (blue line). When we degrade the prediction skill (the experiment 2.2; see Table 4), forecasters still maintain the same level of relative loss by setting low (or zero) predefined warning thresholds issuing many false alarms (orange line). It is apparently unrealistic. In the framework of the GL model, this unrealistic model's behavior can be eliminated by setting the high cost of the mitigation and protection action responding to the issued warning. When we assume the high cost of preparedness actions (the experiment 2.3; see Table 4), the small predefined warning threshold induces high relative loss (green line). Forecasters need to avoid issuing false alarms when the cost which should be paid with false alarms is large.

The SKK model can give different explanation of the avoidance of false alarms. Figure 2b shows the relationship between relative loss and predefined warning thresholds simulated by the SKK model in the experiment 2 (see Table 4). Although we assumed no cost and an accurate prediction system (the experiment 2.4; see Table 4), forecasters need to avoid issuing false alarms by the relatively high predefined warning thresholds to minimize relative loss (blue line). Due to the cry wolf effect found in Figure 1b, forecasters need to decrease the number of false alarms to mitigate the damage of flooding even if there were no cost of false alarms. In other words, forecasters in the SKK model need to pay "implicit cost" of false alarms because false alarms induce not only the cost of mitigation and protection actions for nothing at the current time but also the increase of damages of the future floods by reducing the social collective trust and preparedness.

When we degrade the prediction accuracy (the experiment 2.5; see Table 4), relative loss is more sensitive to predefined warning thresholds (orange line) because the selection of the threshold is more important to accurately detect flood events and reduce the number of false alarms when the prediction is more inaccurate and uncertain. When we consider the high cost of mitigation and protection actions (the experiment 2.6; see Table 4), small predefined warning thresholds further increase relative loss (green line).

Figure 3a compares the GL and SKK models in the green society. In the previous experiments 1 and 2, the damage threshold, $\delta$, is set to 0.35, which is same as Girons Lopez et al. (2017). In the experiments 3.1 and 3.2 (see Table 5), the damage threshold is reduced to 0.20, so that the number of flood events increases. In this case, the GL and SKK models behave similarly. Figure 3c shows time-averaged social collective memory, social collective trust in FEWS, and social preparedness as functions of predefined warning thresholds. In the green society, frequent flood events make social collective memory high. In addition, it is easy to maintain the high social collective trust in FEWS since there are many opportunities to gain trust when flood frequently occurs. Therefore, both social collective memory and social collective trust in FEWS are large in the green society. Although the GL model neglect the social collective trust in FEWS to calculate social preparedness, the social preparedness of both GL and SKK models is high

On the other hand, the GL and SKK models work more differently in the technological society than the green society. The damage threshold, $\delta$, is increased to 0.45 in the experiments 3.3 and 3.4 (see Table 5), so that the number of flood events is smaller than Girons Lopez et al. (2017). Figure 3b indicates that the relationship between relative loss and predefined warning thresholds in the GL model is substantially different from that in the SKK model. The SKK model produces smaller relative loss than the GL model when the appropriate predefined warning threshold is chosen. The sensitivity of relative loss to predefined warning thresholds is larger in the technological society than the green society. Figure 3d indicates that it is difficult to maintain the high level of social collective memory in the technological society, so that considering social collective trust in FEWS can increase social preparedness. In addition, the choice of a predefined warning threshold is more important to maintain the high level of social collective trust in the technological society than the green society.

In the experiment 4, we further analyze the SKK model to discuss the optimal predefined warning threshold and to provide the useful implication for the design of FEWS in the various kind of social systems. We have three sets of parameters in equation (9) (see also Table 6). The first set of parameters is same as the experiments 1-3. Changes in social collective trust by false negative and false positive are same ($\tau_{FN} = \tau_{FP}$). In the second set of parameters, we assume social collective trust substantially decreases by false positive (false alarms) ($\tau_{FN} < \tau_{FP}$): $[\tau_{TP}, \tau_{FN}, \tau_{FP}] = [0.1, 0.1, 0.8]$. In the third set of parameters, we assume social collective trust substantially decreases when forecasters miss a flood event ($\tau_{FN} > \tau_{FP}$): $[\tau_{TP}, \tau_{FN}, \tau_{FP}] = [0.1, 0.8, 0.1]$. The blue, orange, and green lines in Figures 4a-4d show that the optimal predefined warning threshold depends on how social collective trust is affected by false alarms and missed events. When social collective trust is affected by false alarms more substantially than missed events (orange lines), forecasters need to have relatively high predefined warning thresholds to maintain the high level of social collective trust (see Figures 4e-h) and minimize relative loss. Figures 4a-4d also shows that the differences of optimal predefined warning thresholds in three sets of parameters become larger as forecasts become accurate. The optimal predefined thresholds are bounded by the range in which the high threat scores can be obtained (see Figures 4i-4l). Thus, more accurate prediction systems make it more important to change the predefined warning threshold according to the dynamics of social collective trust. It implies that forecasters need to prioritize the meteorologically accurate forecasting by maximizing threat scores. Then, they have a room for improvement to change their warning thresholds based on the dynamics of social collective trust in FEWS.

## 5. Discussion and conclusions

In this study, we included the dynamics of social collective trust in FEWS into the existing socio-hydrological model. By formulating social preparedness as a function of social collective trust as well as social collective memory, we realistically simulate the cry wolf effect, in which many false alarms undermine the credibility of the early warning systems. Please note that the previous version of the model proposed by Girons Lopez et al. (2017) cannot do it. Although our model is simple and stylized, we can provide useful implication to improve the design of FEWS. First, considering the dynamics of social collective trust in FEWS is more important in the technological society with infrequent flood events than in the green society with frequent flood events. Second, as the natural scientific skill to predict flood is improved, the efficiency of FEWS gets more sensitive to the behavior of social collective trust, so that forecasters need to determine their forecasting threshold by considering the social aspects.

Although our model is the small extension of Girons Lopez et al. (2017), the implication of our study is completely different from Girons Lopez et al. (2017). Girons Lopez et al. (2017) mainly focused on the influence of the recency of flood experience on social preparedness and the efficiency of FEWS. Since their social preparedness is determined only by the flood experiences and they did not consider social collective trust in FEWS and weather agencies, the outcome of prediction did not directly influence the people's behavior in the model of Girons Lopez et al. (2017). By formulating social preparedness as a function of both social collective memory and trust, we could evaluate the effects of missed events and false alarms on preparedness actions. We contributed to connecting the modeling approaches of system dynamics in socio-hydrology to the existing literature about complex human behaviors against disaster warnings such as cry wolf effects in economics, sociology, and psychology (e.g., Simmons and Sutter 2009; Ripberger et al. 2015; Trainor et al. 2015; LeClerc and Joslyn 2015; Jauernic and van den Broeke 2017; Lim et al. 2019)

Our findings of the optimal predefined warning thresholds are similar to Roulston and Smith (2003). Roulston and Smith (2003) developed the simple model to optimize predefined warning thresholds considering the damage, cost, and imperfect compliance with forecasting (i.e., the cry wolf effect). They also revealed that it is necessary to choose high warning thresholds if intolerance of false alarms of the society is high. However, there are substantial differences between our study and the previous cost-loss analysis such as Roulston and Smith (2003). First, Roulston and Smith (2003) developed the static model in which the cry wolf effect is treated exogenously while our model is the dynamic model in which the cry wolf effect is endogenously simulated. Therefore, our model can consider the temporal change in the design and accuracy of FEWS, the flood protection level, and social systems, which may be the significant advantage to analyze the actual socio-hydrological phenomena. Second, by fully utilizing the previous achievements of Girons Lopez et al. (2017), we can also consider social collective memory of past disasters, which is not considered by Roulston and Smith (2003). This feature of our model can reveal that the social collective memory also contributes to the optimal predefined warning thresholds.

The major limitation of this study is that our modeling of social collective trust is simple and is not fully supported by empirical data. Although intuition and theory suggest that many false alarms reduce the preparedness actions responding to warnings, the existence of the cry wolf effect in the weather-related disasters is still debatable (see a comprehensive review of Lim et al. (2019)). Simmons and Sutter (2009) indicated that the recent false alarms negatively impacted the preparedness actions, so that we modeled the change in social collective trust by the recent forecast outcome. However, Ripberger et al. (2015) could not find the statistically significant short-term effect of false alarms although they found the statistically significant cry wolf effect using the long-term data. It should be noted that most of previous studies related to the cry wolf effect focused on tornado disasters and the systematic econometric analyses have not been implemented for flood disasters. The effect of social collective memory on catastrophic disasters in the actual society is also debatable (e.g., Fanta et al. 2019). As Mostert (2018) suggested, it is crucially important to perform case study analyses, obtain empirical data, and integrate those data into the dynamic model to deepen our understanding of the hypothesis of the models (e.g., Roobavannan et al. 2017; Ciullo et al. 2017; Barendrecht et al. 2019; Sawada and Hanazaki 2020).

In socio-hydrology, researchers have mainly focused on the functions of land use change and water-related infrastructures such as dams, levees, and dikes in the complex social systems. Although the interactions between social systems and weather forecasting such as the cry wolf effect are interesting, the function of FEWS and weather-related disaster forecasting has not been intensively investigated in socio-hydrology. We call for the new research regime, socio-meteorology, as extension of socio-hydrology. In socio-meteorology, researchers may focus on how social systems interact with water-related disaster forecasting, how the efficiency of weather forecasting is affected by the other hydrological factors such as land use and flood protection infrastructures, and how weather forecasting affects the design of land use and flood protection infrastructures.

**Acknowledgements**

We used the source code of Girons Lopez et al. (2017) which can be downloaded at https://github.com/GironsLopez/prep-fews. This study does not contain any data. This study was supported by the JST FOREST program (grant no. JPMJFR205Q).

Table 1. Summary of the outcomes of the flood early warning system. Loss by each outcome is also shown (see also Section 2).

|             | $Q < \delta$            | $Q \geq \delta$              |
|-------------|-------------------------|------------------------------|
| $P < \pi$   | True negative: 0        | False negative: $D_Q$        |
| $P \geq \pi$| False positive: $C$     | True positive: $C + D_r$     |

Table 2. Fixed model parameters

| | description | equation | values |
|---|---|---|---|
| $\kappa_c$ | shape of the bivariate gamma distribution to generate river discharge timeseries | (1) | 2.5 |
| $\theta_c$ | scale of the bivariate gamma distribution to generate river discharge timeseries | (1) | 0.08 |
| $\mu_m$ | mean of prediction error | (2) | 0 |
| $\beta$ | parameter of the damage function | (3) | 0.2 |
| $\alpha_0$ | minimum residual damage fraction | (4) | 0.2 |
| $\lambda$ | social collective memory decay rate | (8) | 0.028 |
| $\chi$ | psychological shock magnitude | (8) | 1.0 |

**Table 3.** Model parameters in the experiment 1.

|  | description | equation | values |
|---|---|---|---|
| $\sigma_m$ | standard deviation of prediction error | (2) | 0.075 |
| $\mu_v$ | mean of prediction precision | (2) | 0.15 |
| $\sigma_v$ | standard deviation of prediction precision | (2) | 0.075 |
| $\delta$ | Damage threshold | (3,5) | 0.35 |
| $\eta$ | cost parameter | (6) | 0.02 |
| $\gamma$ | Parameter controlling weights of social collective memory and trust | (7) | 1 (GL model) <br> 0.5 (SKK model) |
| $\tau_{TP}$ | Increment of trust for true positive | (9) | 0.1 |
| $\tau_{FN}$ | Increment of trust for false negative | (9) | 0.1 |
| $\tau_{FP}$ | Increment of trust for false positive | (9) | 0.1 |

Table 4. Model parameters in the experiment 2

|  | description | equation | values | | | | | |
|---|---|---|---|---|---|---|---|---|
|  |  |  | exp2.1 | exp2.2 | exp2.3 | exp2.4 | exp2.5 | exp2.6 |
| $\sigma_m$ | standard deviation of prediction error | (2) | 0.05 | 0.075 | 0.05 | 0.05 | 0.075 | 0.05 |
| $\mu_v$ | mean of prediction precision | (2) | 0.05 | 0.15 | 0.05 | 0.05 | 0.15 | 0.05 |
| $\sigma_v$ | standard deviation of prediction precision | (2) | 0.025 | 0.075 | 0.025 | 0.05 | 0.075 | 0.025 |
| $\delta$ | Damage threshold | (3,5) | 0.35 | 0.35 | 0.35 | 0.35 | 0.35 | 0.35 |
| $\eta$ | cost parameter | (6) | 0 | 0 | 0.1 | 0 | 0 | 0.1 |
| $\gamma$ | Parameter controlling weights of social collective memory and trust | (7) | 1 (GL model) | 1 (GL model) | 1 (GL model) | 0.5 (SKK model) | 0.5 (SKK model) | 0.5 (SKK model) |
| $\tau_{TP}$ | Increment of trust for true positive | (9) | 0.1 | 0.1 | 0.1 | 0.1 | 0.1 | 0.1 |
| $\tau_{FN}$ | Increment of trust for false negative | (9) | 0.1 | 0.1 | 0.1 | 0.1 | 0.1 | 0.1 |
| $\tau_{FP}$ | Increment of trust for false positive | (9) | 0.1 | 0.1 | 0.1 | 0.1 | 0.1 | 0.1 |

Table 5. Model parameters in the experiment 3

| | description | equation | values | | | |
|---|---|---|---|---|---|---|
| | | | exp3.1 | exp3.2 | exp3.3 | exp3.4 |
| $\sigma_m$ | standard deviation of prediction error | (2) | 0.05 | 0.05 | 0.05 | 0.05 |
| $\mu_v$ | mean of prediction precision | (2) | 0.05 | 0.05 | 0.05 | 0.05 |
| $\sigma_v$ | standard deviation of prediction precision | (2) | 0.025 | 0.025 | 0.025 | 0.025 |
| $\delta$ | Damage threshold | (3,5) | 0.20 | 0.20 | 0.45 | 0.45 |
| $\eta$ | cost parameter | (6) | 0.02 | 0.02 | 0.02 | 0.02 |
| $\gamma$ | Parameter controlling weights of social collective memory and trust | (7) | 1 (GL model) | 0.5 (SKK model) | 1 (GL model) | 0.5 (SKK model) |
| $\tau_{TP}$ | Increment of trust for true positive | (9) | 0.1 | 0.1 | 0.1 | 0.1 |
| $\tau_{FN}$ | Increment of trust for false negative | (9) | 0.1 | 0.1 | 0.1 | 0.1 |
| $\tau_{FP}$ | Increment of trust for false positive | (9) | 0.1 | 0.1 | 0.1 | 0.1 |

Table 6. Model parameters in the experiment 4.

| | description | equation | values |
|---|---|---|---|
| $\sigma_m$ | standard deviation of prediction error | (2) | 0.05 (accurate forecast) <br> 0.075 (inaccurate forecast) |
| $\mu_v$ | mean of prediction precision | (2) | 0.05 (accurate forecast) <br> 0.15 (inaccurate forecast) |
| $\sigma_v$ | standard deviation of prediction precision | (2) | 0.025 (accurate forecast) <br> 0.075 (inaccurate forecast) |
| $\delta$ | Damage threshold | (3,5) | 0.20 (green society) <br> 0.45 (technological society) |
| $\eta$ | cost parameter | (6) | 0.02 |
| $\gamma$ | Parameter controlling weights of social collective memory and trust | (7) | 1 (GL model) |
| $[\tau_{TP}, \tau_{FN}, \tau_{FP}]$ | Increment of trust for true positive, false negative, and false positive | (9) | [0.1, 0.1, 0.1] (blue lines in Figures 4a-4h) <br> [0.1, 0.1, 0.8] (orange lines in Figures 4a-4h) <br> [0.1, 0.8, 0.1] (green lines in Figures 4a-4h) |

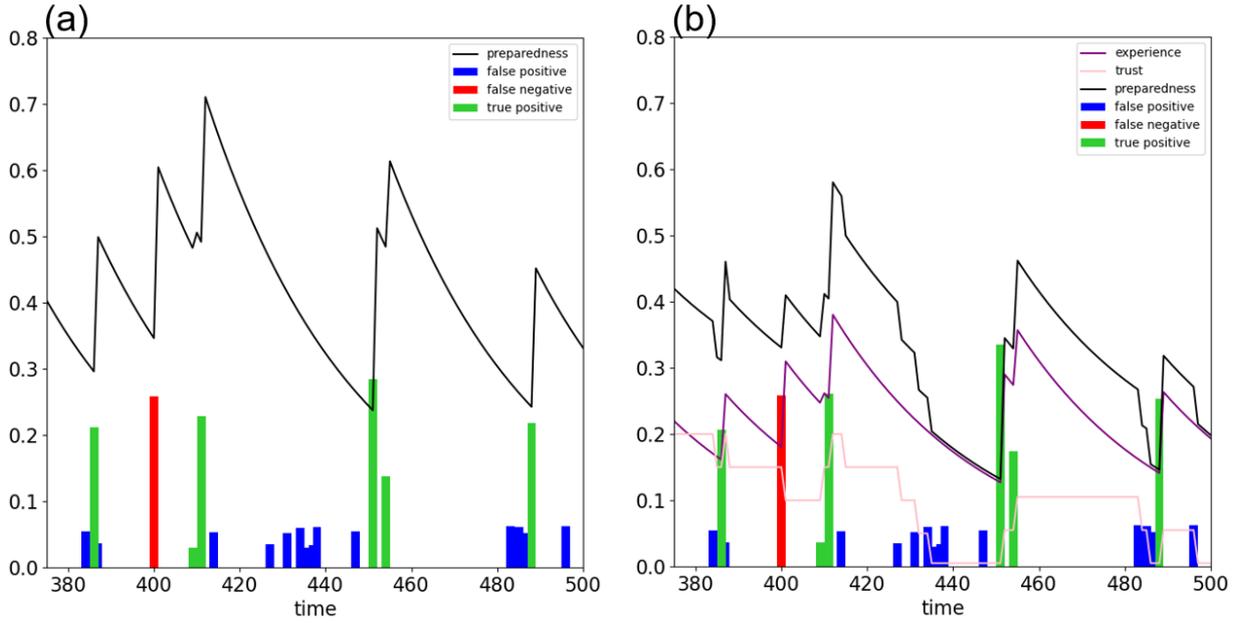

**Figure 1.** Timeseries of (a) the GL model and (b) the SKK model of the experiment 1 (see section 3 and Table 2 for model parameters). Black, purple, and pink lines are social preparedness, half of social collective memory, and half of social collective trust in FEWS, respectively. Since social preparedness is identical to social collective memory and social collective trust is not considered in the GL model, there are no purple and pink lines in (a). Blue, red, and green bars show total loss by the outcomes of false positive, false negative, and true positive, respectively.

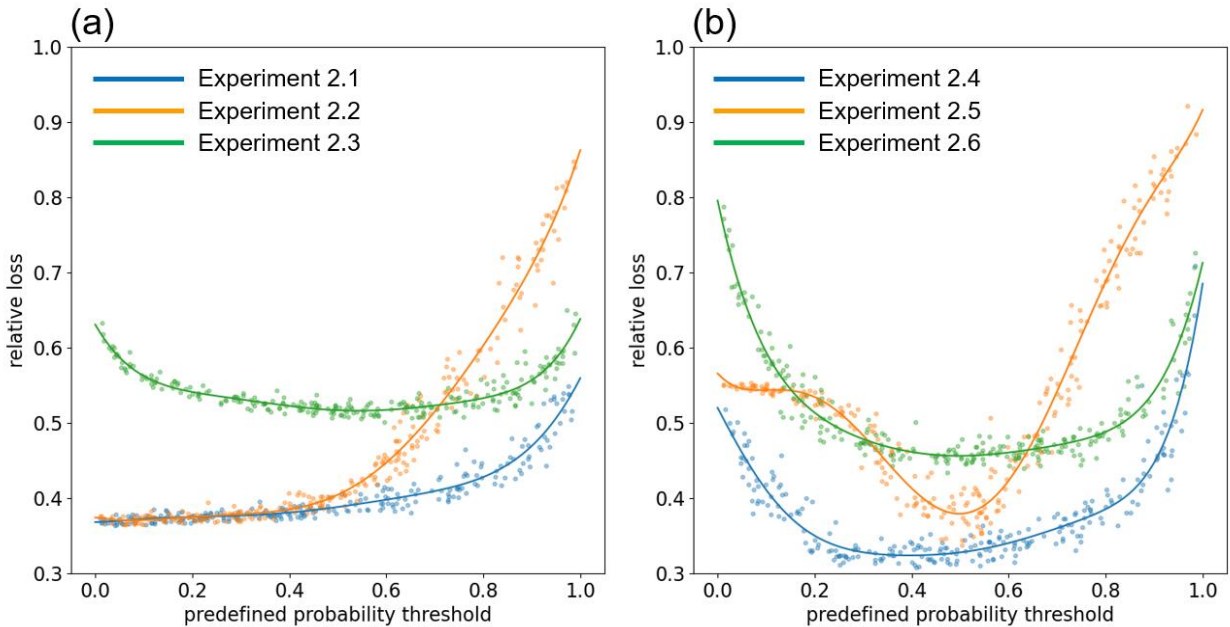

**Figure 2.** The relationship between relative loss and predefined warning thresholds in (a) the GL model and (b) the SKK model in the experiment 2. In (a), blue, orange, and green lines show the results of the experiments 2.1, 2.2, 2.3, respectively. In (b), blue, orange, and green lines show the results of the experiments 2.4, 2.5, 2.6, respectively. Each dot shows the result of the individual Monte-Carlo simulation and we smoothed them by Gaussian process regression. See also Table 4 for detailed parameter settings.

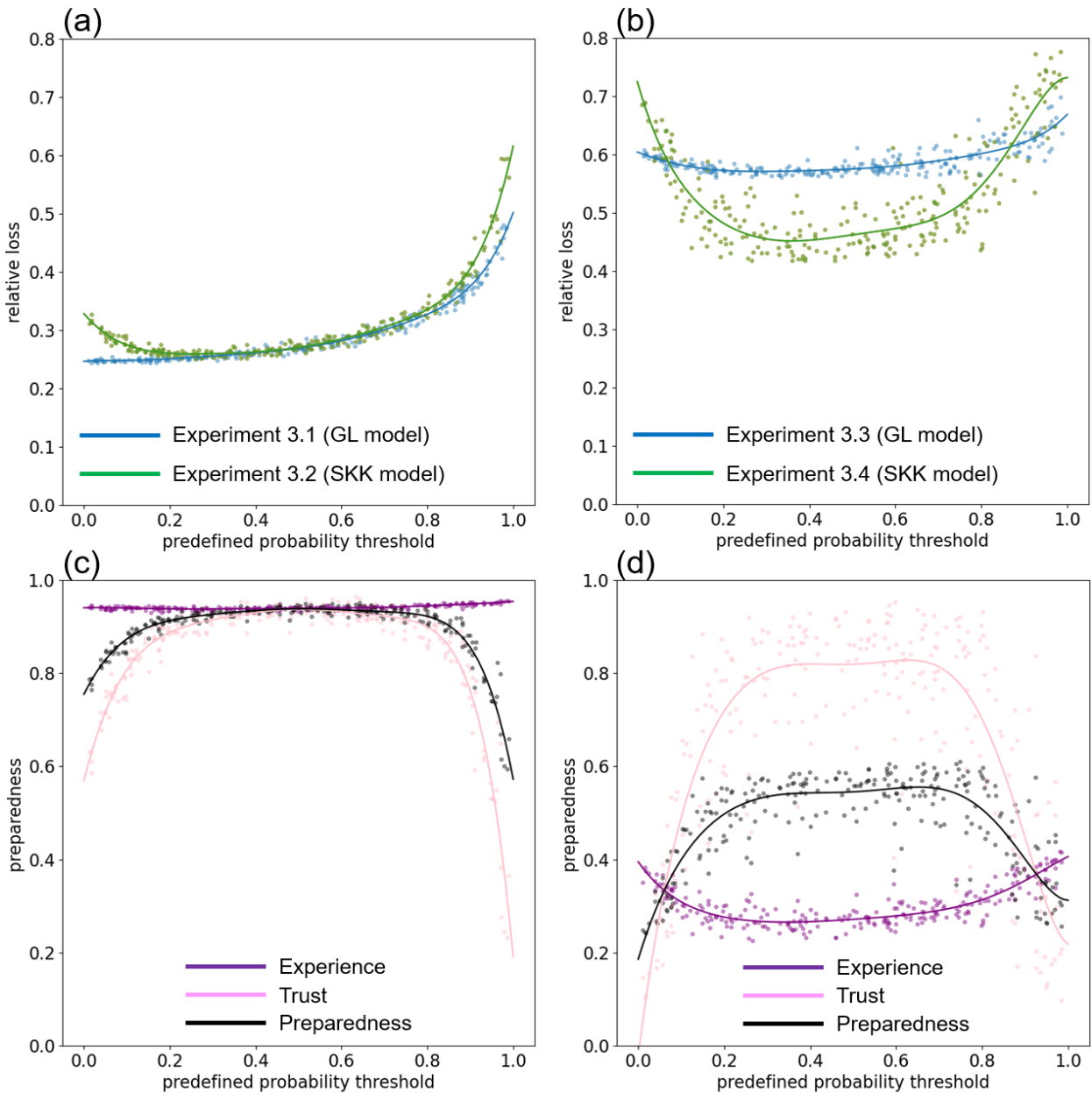

**Figure 3**. (a-b) The relationship between relative loss and predefined warning thresholds in (a) the green society and (b) the technological society. In (a), blue and green lines show the results of the experiments 3.1 and 3.2, respectively. In (b), blue and green lines show the results of the experiments 3.3 and 3.4, respectively. (c-d) The relationship between time-averaged social preparedness and predefined warning thresholds in (c) the green society and (d) the technological society. Black, purple, and pink lines show time-averaged social preparedness, social collective memory, and social collective trust in FEWS. Each dot shows the result of the individual Monte-Carlo simulation and we smoothed them by Gaussian process regression.

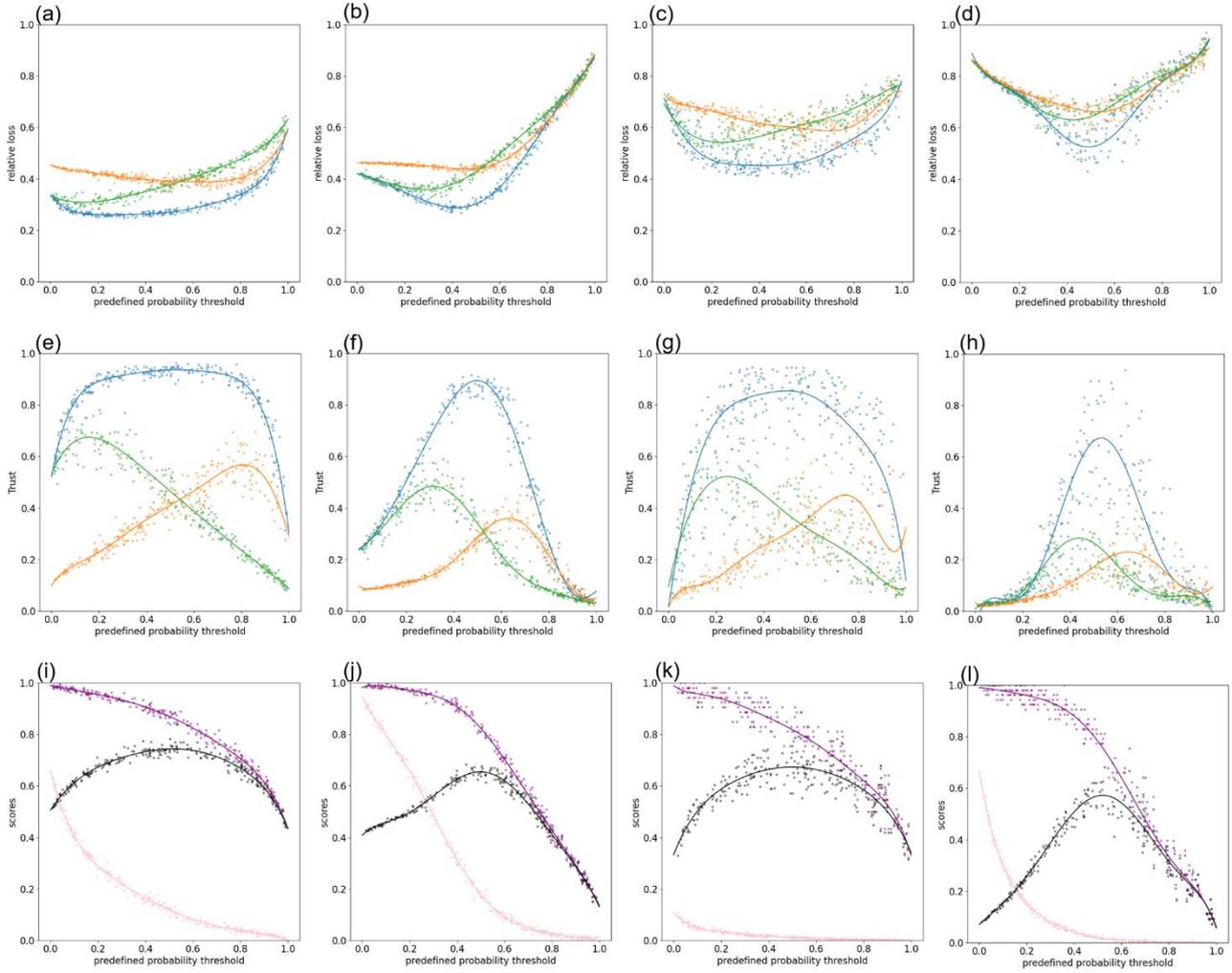

**Figure 4.** Results of the experiment 4. (a-d) The relationship between relative loss and predefined warning thresholds in (a) the green society with accurate forecasts, (b) the green society with inaccurate forecasts, (c) the technological society with accurate forecasts, (d) the technological society with inaccurate forecasts. Increments of trust for true positive, false negative, and false positive are set to 0.1, 0.1, and 0.1 (blue lines), 0.1, 0.1, and 0.8 (orange lines), and 0.1, 0.8, and 0.1 (green lines). See Table 6 for detailed model parameters' settings. (e-f) Same as (a-d) but for time-averaged social collective trust in FEWS. (i-l) Same as (a-d) but for threat score (black lines), hit rate (purple lines), and false alarm ratio (pink lines). Each dot shows the result of individual Monte-Carlo simulation and we smoothed them by Gaussian process regression.